\documentclass[runningheads]{svmult}

\usepackage{epsfig}
\usepackage{makeidx}   
\usepackage{graphicx}  
\usepackage{subeqnar}  
\usepackage{multicol}  
\usepackage{physmubb}  
\makeindex             

\def\lapprox{\lower .7ex\hbox{$\;\stackrel{\textstyle <}{\sim}\;$}}
\def\gapprox{\lower .7ex\hbox{$\;\stackrel{\textstyle >}{\sim}\;$}}

\begin{document}
\title*{Theoretical Developments on Hard QCD Processes at Colliders
   \thanks{Invited talk
     at the \emph{14th Topical Conference on Hadron Collider
       Physics} (HCP2002), Karlsruhe, Germany, 30 Sep--4 Oct 2002}
}
\toctitle{Theoretical Developments on Hard QCD Processes at Colliders}
%
%
\titlerunning{Hard QCD at Colliders}
%
\author{Thomas Gehrmann\\[2mm]{\it Institut 
f\"ur Theoretische Physik,
RWTH Aachen, D-52056 Aachen
}}

\authorrunning{Thomas Gehrmann}
%
%

\maketitle              

\vspace{-6mm}
\section{Introduction}
Quantum Chromodynamics (QCD) is well established as theory of strong 
interactions through a large number of experimental verifications. 
The era of `testing QCD' is clearly finished,
and QCD today is becoming precision physics. In contrast to LEP and 
SLC, where electroweak reactions could be studied to a highly precise 
level without having to take QCD effects into account, QCD is 
ubiquitous at hadron colliders, affecting all observables. 
Any precision measurement (strong coupling constant, quark masses, 
electroweak parameters, parton distributions) 
at the Tevatron and the LHC, as well as any prediction of new 
physics effects and their backgrounds, relies on the understanding of 
QCD effects on the observable under consideration. 

Precision QCD poses several severe challenges from the theoretical 
and computational point of view. Most importantly, the strong coupling 
constant is considerably larger than the electromagnetic coupling constant 
at scales typically probed at colliders: $\alpha_s(M_Z) \simeq 15 
\alpha_{{\rm em}} (M_Z)$, resulting in a slower convergence of the 
perturbative expansion. As a consequence, a precise description of 
QCD observables (precise means here that the theoretical 
uncertainty becomes similar to the achieved or projected experimental errors)
 is obtained only by including higher order corrections, often 
requiring beyond the next-to-leading order. Another important 
challenge is the fact that QCD describes quarks and gluons, while 
experiments observe hadrons. This mismatch is either accounted for by 
a description of the parton to hadron 
transition through fragmentation functions or by defining sufficiently 
inclusive final state observables, such as jets. 
Finally, many collider observables involve largely different scales, such as 
quark masses, transverse momenta and vector boson masses. These give rise 
to potentially large logarithms, which might spoil the convergence of the 
perturbative series and need to be resummed to all orders. 

In this talk, I shall try to highlight recent theoretical progress towards 
precision QCD at colliders, focusing on heavy quark production in 
Section~\ref{sec:hq}, on jets and multiparton final states in 
Section~\ref{sec:jets} and on photons and electroweak bosons in 
Section~\ref{sec:vecbos}. Finally, a summary of the current state-of-the-art
and of yet open issues is given in Section~\ref{sec:conc}.

\section{Heavy Quarks}
\label{sec:hq}
The production of heavy quarks is one of the main topics investigated at 
high energy collider experiments. Heavy quarks are of particular interest 
to elucidate the flavour sector of the standard model, which is less well 
tested than the gauge sector. 
Also, many approaches to physics beyond the standard model, often 
related to electroweak symmetry breaking and mass generation, predict 
new effects to be most pronounced in observables involving heavy quarks. 
\begin{figure}[b]
\begin{center}
\epsfig{file=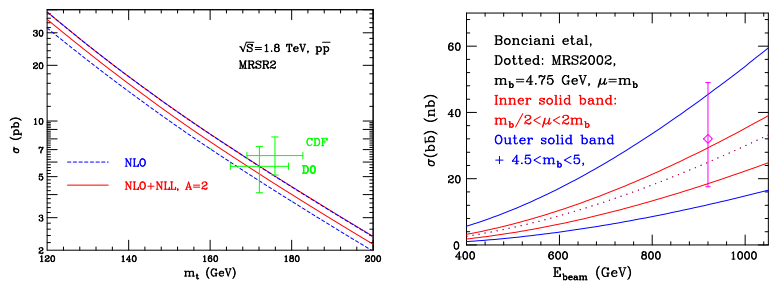,width=5cm}\hspace{4mm} 
\epsfig{file=FT-mrs-f.eps,width=4.63cm} 
\end{center}
\caption{Total cross sections for $t\bar t$ at the Tevatron and 
$b\bar b$ at HERA-B.}
\label{fig:hqtotal}
\end{figure}

\subsection{Total Cross Sections}
The total cross sections for the production of heavy quarks can be
computed within perturbation theory. The current state-of-the-art is 
a next-to-leading order (NLO) calculation~\cite{hqnlo}, which is
further improved by summing large logarithms due to soft gluon
emission up to the next-to-leading logarithmic (NLL)
level~\cite{hqnll}. As can be seen from Figure~\ref{fig:hqtotal},
these predictions are in good agreement with experimental data on the
total $t\bar t$ cross section at the Tevatron~\cite{hqcoll} 
and the total $b\bar b$ 
cross section at HERA-B~\cite{hqherab}
(which both actually refer to similar kinematical 
values of $m_Q/\sqrt{s}$). The theoretical uncertainty on the
prediction for HERA-B is larger for two reasons: the larger value of 
the strong
coupling at $m_b$ than at $m_t$ and the dominance of $gg$ initial
states in $pN$ collisions (HERA-B) compared to $q\bar q$ dominance in 
$p\bar p$ collisions (Tevatron).

\subsection{Transverse Momentum Distributions}
Differential distributions of hadrons containing $b$ quarks 
measured in hadron-hadron, photon-hadron or photon-photon collisions 
have been in apparent discrepancy with theoretical predictions for 
quite some time. The spectrum of $B^\pm$ hadrons measured at 
CDF~\cite{cdftrmom} 
is one of the most recent examples for this discrepancy.

The theoretical prediction for $B$ meson production involves a convolution 
of the hard matrix element for heavy quark production in parton-parton 
scattering with initial parton distributions and final state fragmentation 
functions describing the non-perturbative transition from a $b$ quark to 
a $B$ hadron. It is in particular the latter which may account for the 
discrepancy between theoretical prediction and experimental data, 
especially since it has been observed~\cite{d0bjet} 
that the transverse momentum 
distribution of $b$-tagged jets~\cite{frixione} (which 
has little sensitivity to fragmentation functions) is in much
better agreement with theoretical predictions. 

The definition of heavy quark
fragmentation functions is not free from ambiguities, since 
some aspects of these functions are actually calculable in 
perturbation theory~\cite{mele}. In extracting these fragmentation functions 
from data on $B$ hadron production in $e^+e^-$ collisions, several choices 
are made, related to the order of perturbation theory, the incorporation of 
mass effects in the matrix elements, the resummation of potentially large
perturbative terms, the correction of data for parton showers or the 
parametric form of the ansatz used in the determination. 
A commonly used parametric form is the so-called Peterson 
ansatz~\cite{peterson}, which contains only a single parameter $\epsilon$,
describing the shape of the non-perturbative fragmentation function. 
Unfortunately, the sensitivity of the fragmentation function
on the assumptions 
used in the extraction from $e^+e^-$ spectra is often overlooked when 
using this fragmentation function to compute heavy hadron spectra 
at colliders. It was pointed out in~\cite{kniehl} that a consistent 
treatment (in this case a purely massless approach) in extracting 
and implementing fragmentation functions can yield a sizable reduction 
of the discrepancy between experimental data and theoretical prediction.

In the kinematical range covered by the heavy hadron production at 
Tevatron, $b$ quark mass effects are sizable~\cite{cacciari1}. An approach
incorporating quark mass effects, perturbatively calculable components of 
the heavy quark fragmentation function~\cite{mele} and resummation of 
large logarithms up to the next-to-leading logarithmic level was 
presented in~\cite{cacciari1} with the fixed-order next-to-leading log
(FONLL) scheme, which requires only a small, genuinely non-perturbative 
component of the fragmentation function to be fitted to $e^+e^-$ data. 
In order to expose the information content actually relevant to 
heavy hadron spectra at hadron colliders, this fit is done in moment 
space. 

\begin{figure}[tb]
\begin{center}
\epsfig{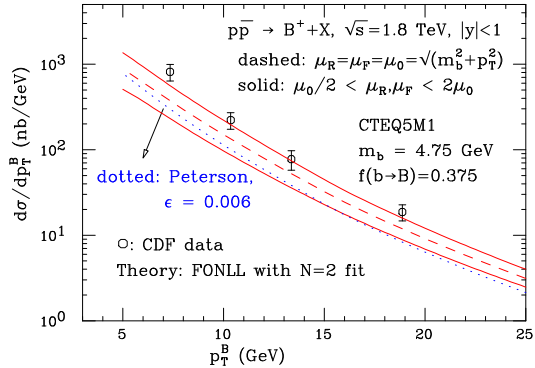}
\end{center}
\caption{Transverse momentum spectrum of $B$ hadrons at CDF, compared 
to calculations using Peterson and FONLL fragmentation functions,
from~\protect\cite{cacciari}.}
\label{fig:hqnll}
\end{figure}

In view of new data from ALEPH~\cite{lephq}, a phenomenological study 
of $B$ hadron production at colliders based on the FONLL 
scheme~\cite{cacciari1} was presented in~\cite{cacciari}. It 
was shown that the consistent treatment of the fragmentation function 
in extraction and prediction reduced the discrepancy between 
data and theoretical prediction considerably, Figure~\ref{fig:hqnll}. 
The theoretical prediction is however still falling somewhat  
short of the experimental data, which is probably due to currently 
uncalculated corrections beyond NLO. 

Many collider experiments also report an excess in the 
$b$ quark production spectra. In interpreting these data, it must 
always be kept in mind that it is not $b$ quark but $B$ hadron production 
which is observed in the experiment. Information on $b$ quark production 
is only inferred from these data using some model for the 
heavy quark fragmentation. As discussed above, there are numerous ambiguities, 
which can yield inconsistent predictions if not implemented consistently.
In view of the rather sizable effects due to a consistent 
treatment of the fragmentation function observed on the $B$ hadron spectra 
at CDF, it might be that the data sets on $b$ quark spectra have to be 
reanalysed incorporating the new experimental information on
 the $b$ quark fragmentation functions in a consistent manner.

\subsection{Top Quark Spin Correlations}
Contrary to $b$ and $c$ quarks, which have lifetimes much longer than 
the typical timescale required for hadronization, $t$ quarks are 
too short-lived to hadronize before they decay. As a consequence,
top quark production and decay can be fully computed within perturbative 
QCD, and the spin state of the $t$ quark determined from the 
distribution of its 
decay products~\cite{ttdecay} 
reflects the $t$ quark spin induced in the production
process, giving rise to non-trivial spin correlations in  $t\bar t$ 
pair production processes~\cite{ttbarhel}. At present, data from 
Tevatron Run I are not yet sufficiently precise to probe these spin 
correlations, results obtained up to now concern only 
the polarization state of the vector boson produced 
in the $t$ decay~\cite{expttbarhel}. Future 
measurements at Run II and LHC  will allow to use $t\bar t$ spin
correlations as a sensitive probe of potential effects beyond the standard 
model showing up in the electroweak vertex of $t$ decay. 

\section{Jets}
\label{sec:jets}
Hadronic jets at large transverse momenta are produced very copiously at 
colliders. Final states with a small number of jets are measured to 
very high experimental accuracy, such that they can be used for 
precision measurements of the strong coupling constant and of 
parton distribution functions. Multiparton final states, involving a 
large number of jets, can on the other hand mimic final state signatures 
introduced by physics beyond the standard model, thus forming an irreducible 
background to searches.
\begin{figure}[tb]
\begin{center}
\epsfig{file=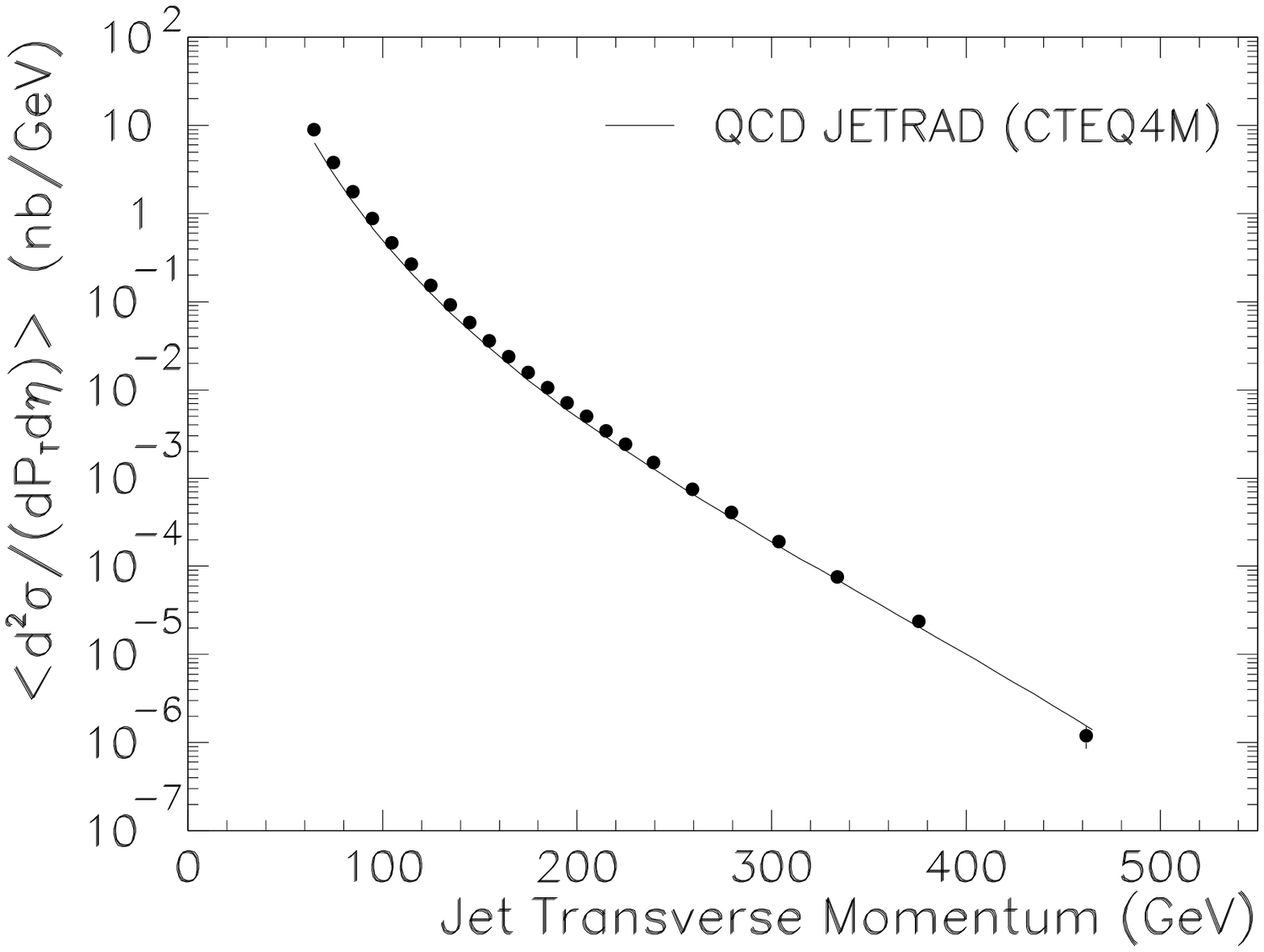,width=6cm} 
\end{center}
\caption{Single jet inclusive cross section at NLO compared to 
D0 data~\protect\cite{d0jet}.}
\label{fig:jetrad}
\begin{center}
\epsfig{file=fig3.epsi,width=6cm} 
\end{center}
\caption{Dependence of the NLO prediction for the $3j$ cross section at the 
Tevatron on renormalization and factorization scale, from \protect\cite{nagy}.}
\label{fig:nlojet}
\end{figure}

The current state of the art for the theoretical description of
jet observables at hadron colliders is next-to-leading order QCD. To 
this order, parton level event generators are available for 
$1j+X$ (EKS \cite{soper} and JETRAD~\cite{jetrad}), 
$2j+X$ (JETRAD \cite{jetrad}), 
$3j+X$ (NLOJET++ \cite{nagy}) and for
$V+X$, $V+1j+X$ (both DYRAD \cite{dyrad}), $V+2j+X$ (MCFM \cite{mcfm}).
In particular, the MCFM project aims to provide a library of a large 
number of final states to NLO accuracy.  
Final states with more than three identified particles (vector bosons or 
jets) are at present only known to leading order. 
NLO predictions do in general yield a very good description of 
the experimental data, e.g.\ for the $1j$ inclusive cross 
section~\cite{d0jet}, 
Figure~\ref{fig:jetrad}. The uncertainty on the theoretical prediction,
as estimated by varying renormalization and factorization scales, is 
reduced considerably from LO to NLO, Figure~\ref{fig:nlojet}.
Nevertheless, for observables such as the single jet inclusive cross section,
the theoretical errors of the NLO prediction are still in excess of the 
experimental errors.

\subsection{Jet Definitions}
Jets in high energy physics experiments are identified by applying a 
jet algorithm to hadronic final states. The default jet algorithm used 
in the analysis of Tevatron Run I data (iterative fixed cone algorithm)
turned out to be unsuited for the study of multi-jet cross sections 
and to be infrared unsafe if a theoretical implementation beyond NLO is 
attempted. These problems can essentially be truncated back to difficulties 
in splitting and merging neighbouring cones, which is done using an 
ad-hoc separation parameter. Several improved algorithms have been suggested,
which can be grouped into two classes: the $k_T$ type algorithm~\cite{ktalg} 
uses a
clustering procedure combining neighbouring preclusters according 
the transverse momentum  of each cluster and to their mutual spatial 
separation. The improved legacy cone algorithm~\cite{ilca} is an 
improved version of the iterative fixed cone algorithm, which avoids 
the problems of the latter.

\subsection{Precision Jet Physics}
\label{sec:jetnnlo}
\begin{figure}[t]
\begin{center}
\epsfig{file=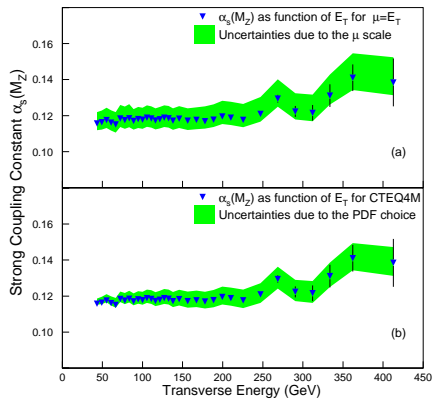,width=6cm} 
\end{center}
\caption{Errors on extraction of $\alpha_s$ from single jet inclusive 
cross section at CDF~\protect\cite{cdfjet}.}
\label{fig:als}
\end{figure}

Despite the evidently good agreement of NLO QCD with experimental data
on jet production rates, Figure~\ref{fig:jetrad}, predictions to this
order are insufficient for many applications. For example, if one uses
data on the single jet inclusive cross section~\cite{cdfjet}
to determine the strong
coupling constant $\alpha_s$, it turns out that the dominant source of 
error on this extraction is due to unknown higher order corrections. 
Given that the theoretical prediction to infinite order in
perturbation theory should be independent of the choice of
renormalization and factorization scale, this error can be estimated
from the variation of the extracted $\alpha_s$ under variation of
these scales, as seen in Figure~\ref{fig:als}. As a result, CDF find 
from their Run I data
\begin{displaymath}
\alpha_s (M_Z)
= 0.1178 \pm 0.0001 (\mbox{stat})^{+0.0081}_{-0.0095}
(\mbox{sys})
\;^{+0.0071}_{-0.0047}(\mbox{scale})\pm 0.0059 (\mbox{pdf}).
\end{displaymath}
It can be seen that the statistical error is already negligible;
improvements in the systematic error can be anticipated in the near
future. To lower the theoretical error, it is mandatory to compute 
next-to-next-to-leading order (NNLO) corrections to the single jet
inclusive cross section.

A similar picture is true in $e^+e^-$ annihilation into three jets and 
deep inelastic $(2+1)$ jet production, where the error on the 
extraction of $\alpha_s$ from experimentally measured jet shape 
observables~\cite{icheplong} is completely dominated by the 
theoretical uncertainty inherent in the NLO QCD calculations. 

Besides lowering the theoretical error, there is  a number of other 
reasons to go beyond NLO in the description of jet
observables~\cite{nigel}. While jets at NLO are modelled theoretically
by at most two partons, NNLO allows up to three partons in a single
jet, thus improving the matching of experimental and theoretical 
jet definitions and resolving the internal jet structure. At hadron
colliders, NNLO does also account for double initial state radiation,
thus providing a perturbative description for the transverse momentum
of the hard final state. Finally, including jet data in a global NNLO
fit of parton distribution functions, one anticipates a lower error on
the prediction of benchmark processes at colliders.

The calculation of jet observables at NNLO requires a number of
different ingredients. To compute the corrections to an $n$-jet
observable, one needs the two-loop $n$ parton matrix elements, 
the one-loop $n+1$ parton matrix elements and the tree level $n+2$
parton matrix elements. Since the latter two contain infrared
singularities due to one or two partons becoming theoretically
unresolved (soft or collinear), one needs to find one- and
two-particle subtraction terms, which account for these singularities 
in the matrix elements, and are sufficiently simple to be integrated
analytically over the unresolved phase space. One-particle
subtraction at tree level is well understood from NLO 
calculations~\cite{ggcs} and general algorithms are available for 
one-particle subtraction at one loop~\cite{onel}. Tree level
two-particle subtraction terms have been studied in the 
literature~\cite{twot},
their integration over the unresolved phase space was up to now made
only in one particular infrared subtraction scheme in the
calculation of higher order corrections to the photon-plus-one-jet
rate in $e^+e^-$ annihilation~\cite{ggam}. A general two-particle subtraction
procedure is still lacking at the moment, although progress on this is 
anticipated in the near future. 

Concerning virtual two-loop corrections to jet-observables related to 
$2\to 2$ scattering and $1\to 3$ decay processes, enormous progress
has been made in the past two years. Much of this progress is due to 
several technical developments concerning the evaluation of two-loop
multi-leg integrals. Using iterative  
algorithms~\cite{laporta}, one can  reduce the
large number of two-loop integrals by means of
integration-by-parts~\cite{chet} and Lorentz invariance~\cite{gr}
identities to a small number of master integrals. The master
integrals relevant to two-loop jet physics are two-loop four-point
functions with all legs on-shell~\cite{onshell} or one leg
off-shell~\cite{mi},
which were computed using explicit integration or implicitly from
their differential equations~\cite{gr}. 
\begin{figure}[t]
\begin{center}
\epsfig{file=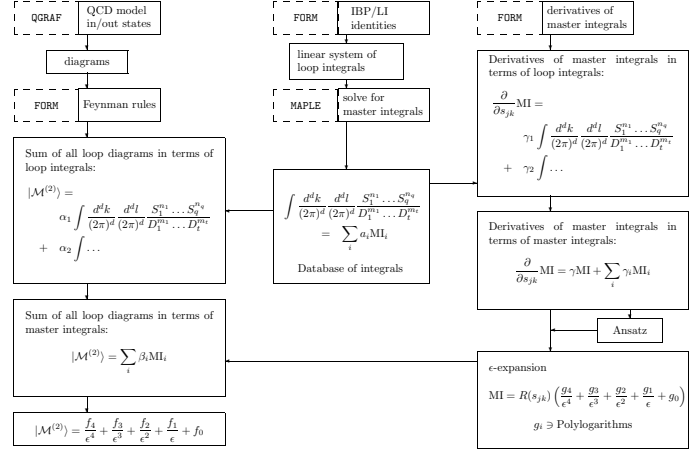,angle=-90,width=9cm} 
\end{center}
\caption{Computer algebra to compute two-loop scattering matrix
  elements for jet physics.}
\label{fig:algebra}
\end{figure}

Combing the reduction scheme with the master integrals, it is 
straightforward to compute the two-loop matrix elements relevant to
jet observables using computer algebra~\cite{radcor}.
The generic structure of such
a calculational procedure is depicted in Figure~\ref{fig:algebra}. 
Following this procedure, two-loop matrix elements were obtained for 
Bhabha scattering~\cite{m1}, parton-parton scattering into two 
partons~\cite{m2}, parton-parton scattering into two 
photons~\cite{m3}, as well as light-by-light
scattering~\cite{m4}. Most recently, two-loop corrections were
computed for the off-shell process $\gamma^*\to q\bar q
g$~\cite{3jme}, relevant to $e^+e^-\to 3j$. Part of these results were
already confirmed~\cite{muw2} using an independent
method~\cite{muw1}. Related to  $e^+e^-\to 3j$ by analytic
continuation~\cite{ancont} are $(2+1)j$ production in $ep$ collisions and $V+j$
production at hadron colliders.
A strong check on all these two-loop results is provided by the
agreement of the singularity structure with predictions obtained from
an infrared factorization formula~\cite{catani}.

\subsection{Multiparton Processes}
Many scenarios for physics beyond the standard model predict
visible signals only in observables with a large number of partons in the 
final state due to production and subsequent decay of 
yet unknown short-lived particles. The same multi-parton final states 
can however also be produced through ordinary QCD processes, acting as
a background to the new physics signals. The extraction of new physics 
effects from multiparton final states does therefore require 
predictions for QCD backgrounds for these final states.  

It turns out that 
a diagrammatic evaluation of the corresponding matrix elements (even using 
modern techniques such as helicity amplitudes) becomes rapidly infeasible due 
to the factorial increase in the number of Feynman diagrams with increasing
number of final state particles. 
An alternative approach~\cite{alpha} circumvents the need for a direct 
evaluation of Feynman diagrams by determining scattering matrix elements 
directly from the numerical Legendre transformation of the interaction 
Lagrangian. This procedure yields only a power-like growth of computation time 
with increasing number of final state particles, and does therefore 
outpace the Feynman diagrammatic evaluation for high multiplicity final 
states. Extending earlier work~\cite{alpha2} on final states with 
up to nine gluons, a parton level event generator ALPHGEN~\cite{alphgen}
to compute 
\begin{itemize}
\item $W/Z/\gamma^* + Q\bar Q + (n\leq 4) j$
\item $H + Q\bar Q + (n\leq 4) j$
\item $W/Z/\gamma^* + (n\leq 6) j$
\item $nW + m Z + l H + N j (n+m+l+N\leq 8, N\leq 3)$
\item $Q\bar Q +  (n\leq 6) j$
\item $Q\bar Q + Q' \bar Q' +  (n\leq 4) j$
\end{itemize}
was presented recently. The event generation provides full colour and flavour 
information, such that a hadronization model and a subsequent detector 
simulation can be interfaced.

Combining matrix element based event generation with parton showers,
special care has to be taken to avoid overcounting of contributions. 
Extending a method derived for $e^+e^-$ annihilation~\cite{krauss1}, 
a procedure for reweighting of events
and determining appropriate starting conditions and cut-off 
of the parton shower has been proposed~\cite{krauss2} 
to allow combination with the 
exact multi-parton matrix elements.

A closely related issue is the incorporation of next-to-leading order 
corrections in parton shower Monte Carlo programs. Recently, 
algorithms were devised for this task by several 
groups~\cite{nlomc1}, first programs 
implementing these algorithms have already been released~\cite{nlomc2}.

\section{Photons and Massive Gauge Bosons}
\label{sec:vecbos}
Photons and gauge bosons provide very prominent final state signatures at 
colliders.
Their study allows the precise determination of electroweak parameters 
at hadron colliders, and their final state signatures are often background
to searches, such as photon pair production to the Higgs search in the 
lower mass range. 

\subsection{Isolated Photons}
Photons produced in hadronic collisions arise essentially from two different 
sources: `direct' or `prompt' photon production via hard partonic processes
such as $qg\to q\gamma$ and $q\bar q\to g\gamma$ or through the `fragmentation'
of a hadronic jet into a single photon carrying a large fraction of the 
jet energy. The former gives rise to perturbatively calculable short-distance 
contributions whereas the latter is primarily a long distance process which 
cannot be calculated perturbatively and is described in terms of the 
quark-to-photon fragmentation function. In principle, this fragmentation 
contribution could be suppressed to a certain extent by imposing isolation 
cuts on the photon. Commonly used isolation cuts are defined by admitting only 
a maximum amount of hadronic energy in a cone of a given radius around the 
photon. An alternative procedure is the democratic clustering approach 
suggested in~\cite{morgan}, which applies standard jet clustering algorithms 
to events with final state photons, treating the photon like any other hadron 
in the clustering procedure. Isolated photons are then defined to be 
photons carrying more than some large, predefined amount of the jet energy. 

Both types of 
isolation criteria infrared safe, although the matching of experimental 
and theoretical implementations of these criteria is in general far from 
trivial. It was pointed out recently~\cite{cf} that cone-based isolation 
criteria fail for small cone sizes $R$ (once $\alpha_s\ln R^{-2}\sim 1$), 
since the isolated photon cross section exceeds the inclusive photon cross 
section. This problem can only be overcome by a resummation of the large 
logarithms induced by the cone size parameter.

The contribution from photon fragmentation to isolated photon 
cross sections at hadron colliders 
is sensitive (for both types of isolation criteria) on 
the photon fragmentation function at large momentum transfer, which 
has up to now been measured only by one of the LEP experiments~\cite{aleph}. 
Based on the democratic clustering procedure, ALEPH extracted the 
quark-to-photon fragmentation function from the measured photon-plus-one-jet
rate. It was observed that the resulting prediction~\cite{morgan} 
for the variation of 
the isolated photon rate with the resolution parameter was in 
good agreement with the measurement, especially once NLO 
corrections~\cite{ggam} were included.
A related OPAL measurement~\cite{opal} 
of the photon fragmentation function from 
inclusive photon production in $e^+e^-$ is unfortunately not sufficiently 
sensitive on the behaviour at large momentum transfers~\cite{ggg}.
Further information on the photon fragmentation function at large momentum 
transfer might be gained from yet unanalysed LEP data or from 
the study of photon-plus-jet final states in deep inelastic scattering at
HERA~\cite{gks}, where first data are now becoming available~\cite{lee}.

\subsection{Photon Pairs}
\begin{figure}[tb]
\begin{center}
\epsfig{file=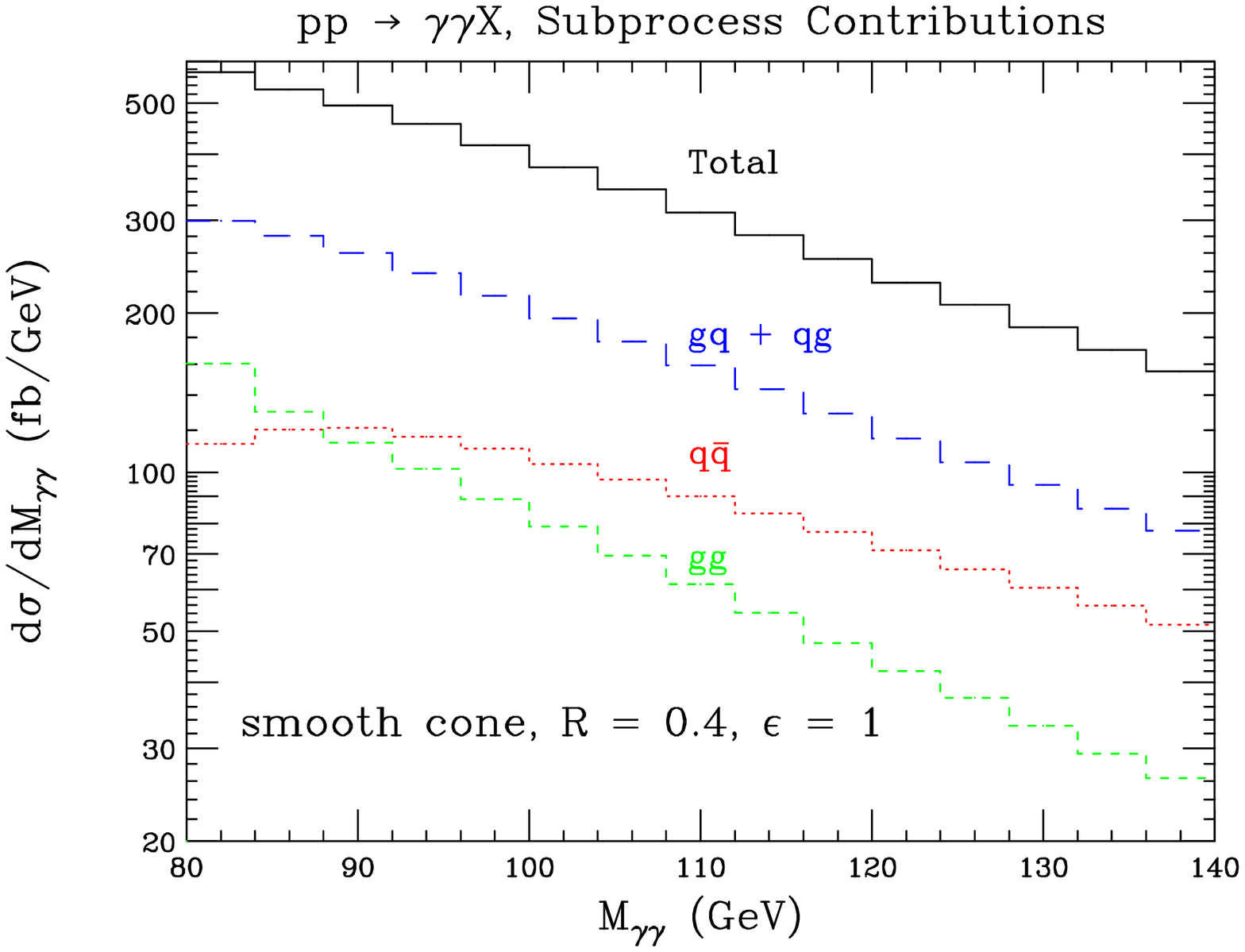,width=5cm}\hspace{4mm}
\epsfig{file=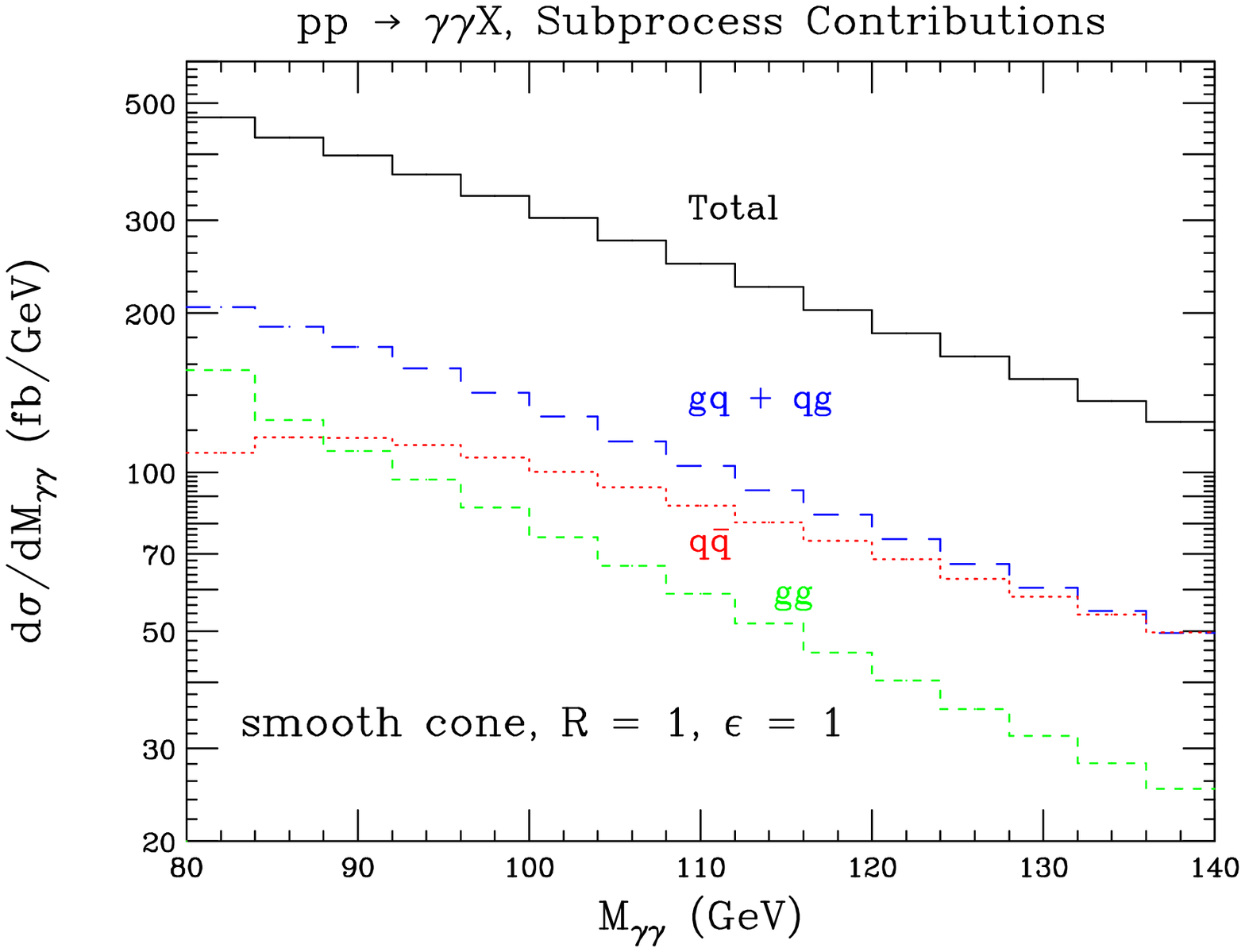,width=5cm}
\end{center}
\caption{Contributions from $q\bar q$, $qg$ and $gg$ subprocesses to 
$\gamma\gamma$ final states
 for different isolation criteria, 
from \protect\cite{bernschmidt}.}
\label{fig:gparton}
\begin{center}
\epsfig{file=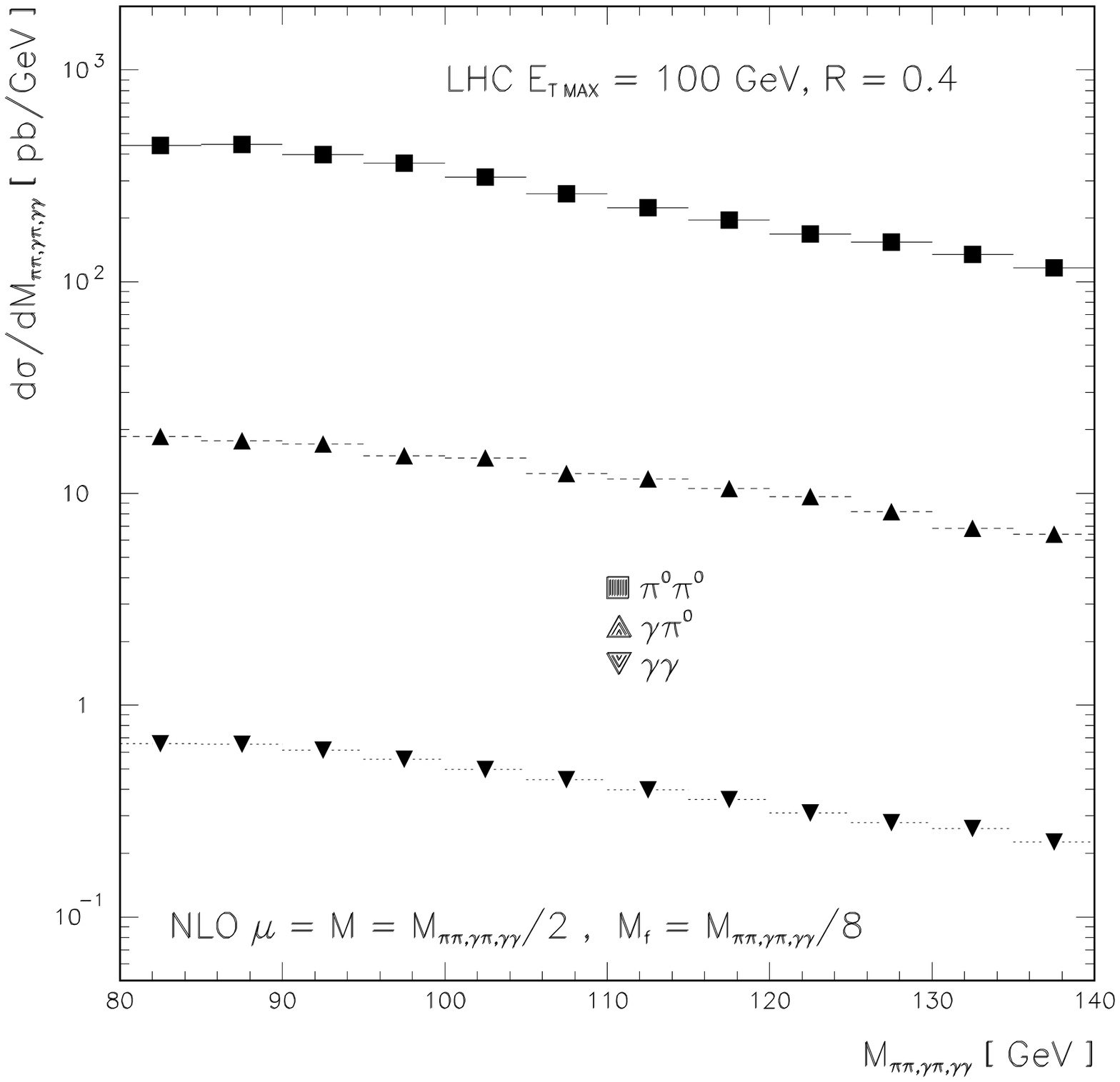,width=5cm}\hspace{4mm}
\epsfig{file=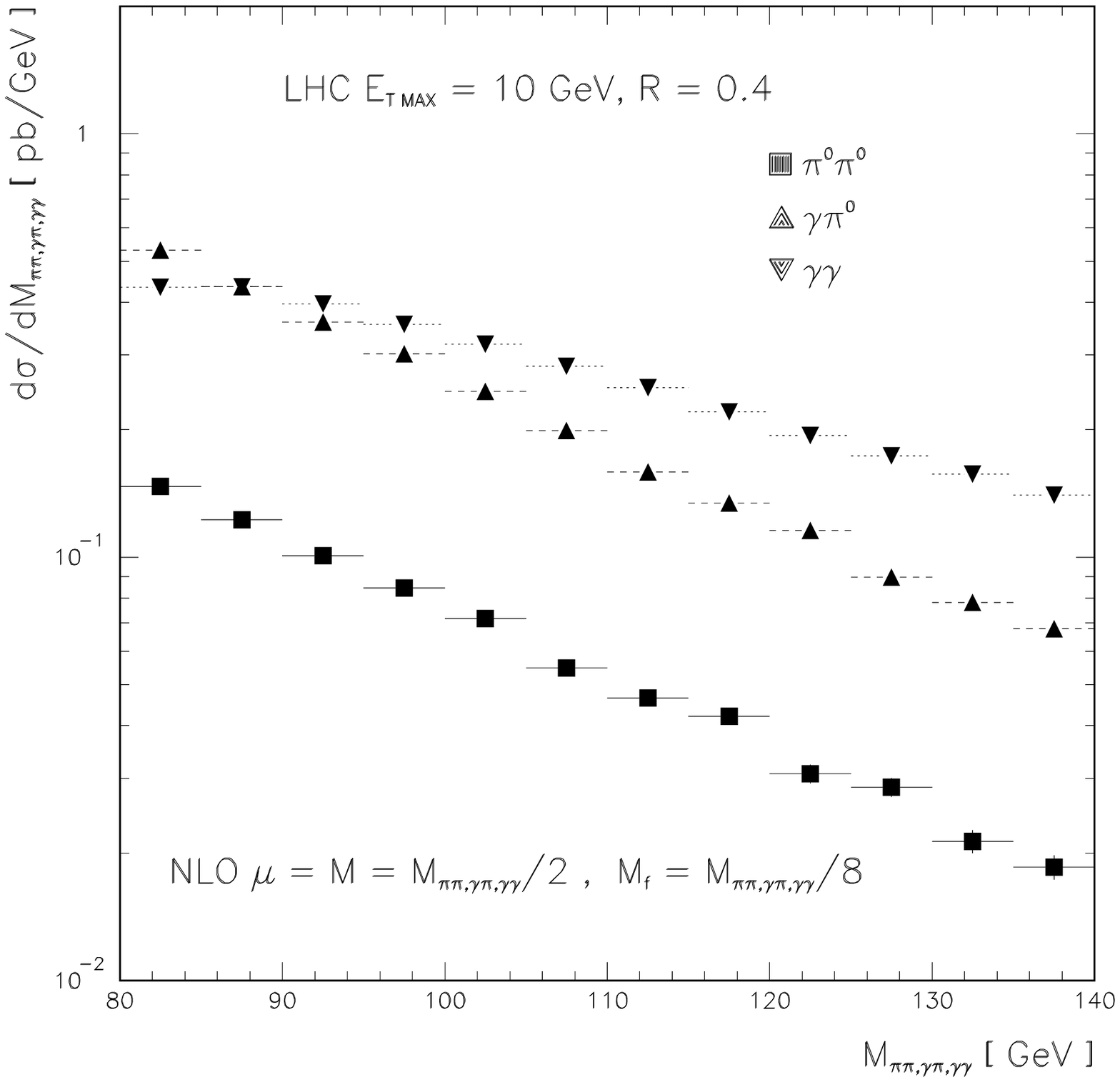,width=5cm}
\end{center}
\caption{Comparison of $\pi^0\pi^0$, $\pi^0\gamma$ and $\gamma\gamma$
production cross sections for different isolation criteria, 
from \protect\cite{diphoxpi}.}
\label{fig:pigamma}
\end{figure}

The discovery of a light Higgs boson ($m_H \lapprox 140$~GeV) at the LHC 
is based largely on the observation of the 
rare decay to two photons~\cite{harlanderhcp}. To perform an accurate 
background subtraction for this observable, one requires a precise 
prediction for QCD reactions yielding di-photon final states. At first 
sight, the ${\cal O}(\alpha_s^0)$ process $q\bar q\to \gamma\gamma$
yields the leading contribution. However, due to the large gluon luminosity 
at the LHC, both $qg\to q \gamma\gamma$ $({\cal O}(\alpha_s^1))$ and 
$gg\to \gamma\gamma$ $({\cal O}(\alpha_s^2))$ 
subprocesses yield contributions 
of comparable magnitude (Figure~\ref{fig:gparton}). The NLO corrections to 
the $q\bar q$ and $qg$ subprocesses have been  known for quite some time,
these are  implemented in the flexible parton level event 
generator DIPHOX~\cite{diphox}. Most recently, NLO corrections were 
also derived for the $gg$ subprocess~\cite{bernschmidt}. Since the 
lowest order contribution to this process is already mediated by a quark loop, 
this calculation contains some of the features appearing in 
jet physics only at NNLO, such as two-loop 
QCD amplitudes and unresolved limits of one-loop amplitudes 
(see Section~\ref{sec:jetnnlo} above).

It must be kept in mind that the di-photon cross sections are highly sensitive 
on the isolation criteria applied to the photons, Figure~\ref{fig:gparton},
with a substantial contribution arising  from photon fragmentation at large 
momentum transfers~\cite{diphoxfrag}. 
Moreover, it is experimentally difficult to distinguish
photons from highly energetic neutral pions which decay into a closely 
collimated photon pair, mimicking a single photon signature.
The pion background in photon pair production has been studied 
to NLO 
in~\cite{diphoxpi} and implemented in DIPHOX, showing that in particular
the $\pi^0\gamma$ channel remains comparable to the $\gamma\gamma$ channel 
even for tight isolation criteria, Figure~\ref{fig:pigamma}. 

\subsection{Vector Boson and Higgs Production}
The production cross sections for $W^\pm$ and $Z^0$ bosons at hadron colliders 
are well understood both experimentally and theoretically. At present,
these cross sections are measured to an error of about 10\% from 
Tevatron Run I, largely limited by statistics. A considerable reduction of the 
experimental error is anticipated from Run II and for the LHC. On the theory
side, inclusive vector boson production has been computed to 
NNLO~\cite{dynnlo} already more than ten years ago.
Very recently, these results have been verified for the first time 
in an independent calculation~\cite{hkhiggs}. The uncertainty on the 
theoretical prediction has been assessed in detail in~\cite{thorne} and
found to be around 3\% from variations of renormalization and factorization 
scale as well as from uncertainties on the parton distribution functions. 
Some improvements are anticipated once full NNLO determinations~\cite{mvv}
of the parton distribution functions become available. 

Given the good theoretical and experimental understanding of 
$W^\pm$ and $Z^0$ boson production, it has been suggested to use these 
for a determination of the LHC luminosity~\cite{dittmar}. In practice, it 
turns out that it is not possible to measure the fully inclusive production 
cross sections, but only cross sections integrated over a  restricted range
in rapidity, which are only known to NLO at 
present~\cite{aem}. For the $W^\pm$ 
production, which is observed only through the $l\nu$ decay channel, it 
is moreover mandatory to compute the spatial distribution of the decay 
products, which is again only known to NLO~\cite{dyrad}. 
To match the anticipated 
experimental accuracy at the LHC and to render vector boson production 
a reliable luminosity monitor, it would be required to extend both these 
calculations to NNLO. 

Closely related to the inclusive vector boson production cross section 
is the inclusive Higgs boson production cross section in the infinite 
top mass limit. 
While the former is induced, at the leading order,
by the partonic subprocess $q\bar q\to W^\pm, Z^0$, the latter is
$gg\to H$.  
The NNLO corrections to inclusive Higgs boson production have been derived,
first in the soft/collinear approximation~\cite{hggsoftcol}; shortly 
thereafter, the full coefficient functions were obtained by 
expansion around the soft limit~\cite{hkhiggs},
 and  fully analytically~\cite{babis} 
by extending the IBP/LI reduction method and the 
differential equation technique (see Section~\ref{sec:jetnnlo}) to
compute double real emission contributions. It turned out that 
inclusion of NNLO corrections yields a sizable enhancement of the Higgs 
production cross section, and a reduction of the uncertainty due to 
renormalization and factorization scale.
These calculations were 
also extended to pseudoscalar Higgs production~\cite{hp}. 
A more detailed discussion of recent results on Higgs physics at colliders 
can be found in~\cite{harlanderhcp}.

\subsection{Transverse Momentum Distributions}
Transverse momentum distributions of vector bosons or any other colour-neutral
final states are well described in perturbation theory~\cite{transmom}
 only if the transverse
momentum $q_T$ is of the same order as the invariant mass $M$
of the colour-neutral
final state. Otherwise, the convergence of the perturbative series is spoilt
by large logarithms $\ln (q_T/M)$. Reliable predictions can be 
obtained if these logarithms are resummed to all orders in perturbation 
theory~\cite{css}. This resummation procedure exponentiates logarithms 
from three different sources: soft $(A)$ and collinear $(B)$ radiation 
associated to the hard interaction (expressed as Sudakov form factor)
and collinear initial state radiation $(C)$. The resummation coefficients
$A,B,C$ can be expanded in powers of the strong coupling constant; it turns 
out that the soft coefficient $A$ only depends on the partonic initial state,
while $B,C$ also depend on the final state under consideration. 

The leading logarithmic (LL) 
corrections are obtained from soft exponentiation 
alone, while collinear contributions  first enter at the next-to-leading
logarithmic (NLL) 
level~\cite{nllresum}. Finally, collinear initial state radiation 
affects only higher (NNLL) order logarithms. At present, 
NNLL resummed
corrections are known for electroweak vector boson 
production~\cite{davies} and Higgs production~\cite{higgsresum}. 
Incorporation of these corrections~\cite{resumincl} improves the theoretical 
description of vector boson production at small transverse momenta 
considerably, and yields a stabilisation of the theoretical prediction 
under variations of renormalization and factorization scales. 

The purely soft $(A)$ terms can be read off from the partonic splitting 
functions to the appropriate order~\cite{mvv} or computed by considering 
only eikonal diagrams~\cite{berger}. 

By reordering the terms in the resummation formula, it is possible to 
extract the final-state independent  terms in the $B$ and $C$ coefficients, and
to group the genuinely process dependent terms into a hard coefficient 
function $H$. This modified resummation formula~\cite{cdfg} allows to 
obtain resummed predictions for a large number of different final states 
(such as vector boson pairs) without having to rederive the full set of 
resummation coefficients.

In the course of the computation~\cite{dfg}
 of these universal resummation coefficients
to NNLL accuracy, one encounters configurations similar to the ones 
appearing in double real radiation contributions to jet physics at NNLO.

Besides the above developments on extending the existing resummation formalism 
towards higher accuracy, considerable progress has been made also to
perform resummation in processes not characterised by a colour 
neutral system with a large invariant mass. In particular, a 
formalism to jointly resum partonic threshold and recoil corrections has been 
developed~\cite{jresum} and applied successfully to prompt photon and 
vector boson production~\cite{jappl}.

\section{Conclusions and Outlook}
\label{sec:conc}
QCD at present and future hadron colliders will be precision physics, 
very much like electroweak physics was precision physics in the LEP era. 
The study of many of the standard scattering reactions will allow a precise
determination of the strong coupling constant, electroweak parameters, 
quark masses and parton distribution functions. In turn, this information 
translates in improved predictions for new physics signals and their 
backgrounds. 

To match the experimental accuracy reached for a number of QCD collider 
observables, the current theoretical predictions have to be improved in 
several aspects. Most importantly, NLO QCD is often not enough if 
confronted with high precision data (such as for jets and gauge bosons)
or observables with slowly converging perturbative expansions (such as
heavy quarks or the Higgs boson). Considerable progress has been made very 
recently in extending QCD calculations to NNLO, first results on 
inclusive observables are now available, and calculations for jet 
observables are well advanced. Many observables do moreover require the 
resummation of large logarithms spoiling the convergence of the perturbative 
series. A universal picture for these resummations started to emerge 
recently, and new techniques have been developed to compute the
corresponding resummation coefficients. Fragmentation effects enter 
many observables with identified particles in the final state. In particular,
a consistent treatment of heavy quark fragmentation effects can account for 
a large part of the observed discrepancy in $B$ hadron spectra, and 
quark-to-photon as well as quark-to-pion fragmentation yield important 
contributions to photon pair final states forming an important
 background to Higgs searches. Much valuable information on these 
fragmentation functions is contained in data from LEP, and 
should be extracted (as long as this is still a feasible task)
to improve predictions for collider observables.

%

\end{document}